\documentclass[10pt,twoside,english,aps,prl,superscriptaddress,twocolumn,notitlepage]{revtex4-1}
\usepackage{mathpazo}

\usepackage[LGR,T1]{fontenc}
\usepackage[latin9]{inputenc}
\usepackage[a4paper]{geometry}
\usepackage{xcolor}
\usepackage{pdfcolmk}
\usepackage{babel}
\usepackage{verbatim}
\usepackage{textcomp}
\usepackage{amsmath}
\usepackage{amssymb}
\usepackage{graphicx}
\PassOptionsToPackage{normalem}{ulem}
\usepackage{ulem}
\usepackage[unicode=true,pdfusetitle,
 bookmarks=false,
 breaklinks=true,pdfborder={0 0 0},pdfborderstyle={},backref=false,colorlinks=true]
 {hyperref}
\hypersetup{
 colorlinks=true,citecolor=black,linkcolor=black,urlcolor=black,pagebackref=true}

\makeatletter

\DeclareRobustCommand{\greektext}{%
  \fontencoding{LGR}\selectfont\def\encodingdefault{LGR}}
\DeclareRobustCommand{\textgreek}[1]{\leavevmode{\greektext #1}}
\ProvideTextCommand{\~}{LGR}[1]{\char126#1}

\providecolor{lyxadded}{rgb}{0,0,1}
\providecolor{lyxdeleted}{rgb}{1,0,0}

\DeclareRobustCommand{\lyxsout}[1]{\ifx\\#1\else\sout{#1}\fi}

\usepackage{amsfonts,anyfontsize,xcolor,graphicx}
\usepackage[calcwidth,explicit]{titlesec}
\usepackage[normalem]{ulem}

\titleformat{\section}{\bfseries\large\sffamily\filcenter}{}{0em}{#1} 
\titlespacing{\section}{0pt}{0.8ex}{0ex}
\titleformat{\subsection}{\bfseries\sffamily\normalsize}{}{0em}{#1} 
\titlespacing{\subsection}{0pt}{0.5ex}{0ex}
\titleformat{\paragraph}[runin]{\bfseries}{}{0pt}{}
\titlespacing*{\paragraph}{0.75em}{0ex}{0.25em}[]

\makeatletter\renewcommand\frontmatter@abstractwidth{\dimexpr\textwidth-2cm\relax}\makeatother

\addto\captionsenglish{}

\setcitestyle{super} 

\makeatletter\@addtoreset{paragraph}{section}\makeatother
\makeatletter\def\p@paragraph{}\makeatother

\AtBeginDocument{
\renewcommand{\ref}[1]{\autoref{#1}}

}

\makeatother

\begin{document}

\title{Gyrotropic resonance of individual Néel skyrmions in Ir/Fe/Co/Pt
multilayers\smallskip{}
}

\author{Bhartendu Satywali}

\thanks{These authors contributed equally to the work.}

\affiliation{Division of Physics and Applied Physics, School of Physical and Mathematical
Sciences, Nanyang Technological University, 637371 Singapore}

\author{Fusheng Ma}

\thanks{These authors contributed equally to the work.}

\affiliation{Division of Physics and Applied Physics, School of Physical and Mathematical
Sciences, Nanyang Technological University, 637371 Singapore}

\author{Shikun He}

\affiliation{Data Storage Institute, Agency for Science, Technology, and Research
(A{*}STAR), 138634 Singapore}

\affiliation{Division of Physics and Applied Physics, School of Physical and Mathematical
Sciences, Nanyang Technological University, 637371 Singapore}

\author{M. Raju}

\affiliation{Division of Physics and Applied Physics, School of Physical and Mathematical
Sciences, Nanyang Technological University, 637371 Singapore}

\author{Volodymyr P. Kravchuk}

\affiliation{Leibniz-Institut für Festkörper- und Werkstoffforschung, IFW Dresden,
D-01171 Dresden, Germany}

\affiliation{Bogolyubov Institute for Theoretical Physics of National Academy
of Sciences of Ukraine, 03680 Kyiv, Ukraine}

\author{Markus Garst}

\affiliation{Institut für Theoretische Physik, TU Dresden, 01062 Dresden, Germany}

\author{Anjan Soumyanarayanan}

\thanks{Correspondence should be addressed to \textbf{A.S. }(\href{mailto:souma@dsi.a-star.edu.sg}{souma@dsi.a-star.edu.sg})
or \textbf{C.P. }(\href{mailto:christos@ntu.edu.sg}{christos@ntu.edu.sg}).\vspace{2ex}}

\affiliation{Data Storage Institute, Agency for Science, Technology, and Research
(A{*}STAR), 138634 Singapore}

\affiliation{Division of Physics and Applied Physics, School of Physical and Mathematical
Sciences, Nanyang Technological University, 637371 Singapore}

\author{C. Panagopoulos}

\thanks{Correspondence should be addressed to \textbf{A.S. }(\href{mailto:souma@dsi.a-star.edu.sg}{souma@dsi.a-star.edu.sg})
or \textbf{C.P. }(\href{mailto:christos@ntu.edu.sg}{christos@ntu.edu.sg}).\vspace{2ex}}

\affiliation{Division of Physics and Applied Physics, School of Physical and Mathematical
Sciences, Nanyang Technological University, 637371 Singapore}
\begin{abstract}
\noindent Magnetic skyrmions are nanoscale spin structures recently
discovered at room temperature (RT) in multilayer films. Employing
their novel topological properties towards exciting technological
prospects requires a mechanistic understanding of the excitation and
relaxation mechanisms governing their stability and dynamics. Here
we report on the magnetization dynamics of RT Néel skyrmions in Ir/Fe/Co/Pt
multilayer films. We observe a ubiquitous excitation mode in the microwave
absorption spectrum, arising from the gyrotropic resonance of topological
skyrmions, that is robust over a wide range of temperatures and sample
compositions. A combination of simulations and analytical calculations
establish that the spectrum is shaped by the interplay of interlayer
and interfacial magnetic interactions unique to multilayers, yielding
skyrmion resonances strongly renormalized to lower frequencies. Our
work provides fundamental spectroscopic insights on the spatiotemporal
dynamics of topological spin structures, and crucial directions towards
their functionalization in nanoscale devices. 
\end{abstract}

\date{February 2018 }

\maketitle

\noindent %

\noindent 
\paragraph{Sk-Intro}

\noindent Magnetic skyrmions are topologically protected spin textures
with solitonic, i.e. particle-like properties\citep{Bogdanov1994,Nagaosa2013}.
Their room temperature (RT) realization in multilayer films possessing
interfacial Dzyaloshinskii-Moriya interaction (DMI)\citep{MoreauLuchaire2016,Woo2016,Jiang2015,Boulle2016,Soumyanarayanan2017}
promises imminent spintronic applications\citep{Finocchio2016,Soumyanarayanan2016},
due to the expected pliability to charge currents and spin torques\citep{Romming2013},
coupled with topological stability\citep{Nagaosa2013,Finocchio2016}.
Harnessing the potential of these magnetic quasiparticles requires
a comprehensive understanding of the excitation and relaxation mechanisms
governing their spatiotemporal dynamics\citep{Garst2017}.%

\paragraph{Sk-Dynamics}

The dynamic response of a skyrmion to electromagnetic potentials includes
a characteristic transverse, or \emph{gyrotropic} component which
derives from its topological charge\citep{Mochizuki2012,Buttner2015},
and is described by distinct rotational modes\citep{Mochizuki2012,Buttner2015,Garst2017}.
Meanwhile, extrinsic perturbations encountered by a moving skyrmion,
e.g. thermal fluctuations or disorder, may deform its spin structure
to an extent governed by internal excitation modes\citep{Mochizuki2012,Lin2014}.
Conversely, similar excitations may be engineered to enable deterministic
skyrmion nucleation in devices\citep{Sampaio2013}. Finally, the efficiency
of its electrodynamic response is governed by spin relaxation and
scattering mechanisms, and characterized by the damping parameter\citep{Finocchio2016}.
A panoramic picture of skyrmion magnetization dynamics in multilayers
is essential for their efficient manipulation in devices utilizing
their mobility in racetracks\citep{Iwasaki2013,Zhou2015}, switching
in dots\citep{Sampaio2013}, and microwave response in detectors and
oscillators\citep{Finocchio2015,Garcia-Sanchez2016}. %

\paragraph{Multilayer Skyrmions}

Skyrmion excitations have thus far been investigated primarily in
crystalline magnets at cryogenic temperatures\citep{Onose2012,Schwarze2015,Ehlers2016,Turgut2017,Montoya2017,Garst2017}.
Such ''helimagnets'' host ordered, dense lattices of identical predominantly
Bloch-textured skyrmions over a narrow temperature range\citep{Nagaosa2013,Schwarze2015}.
In contrast, multilayer skyrmions would likely have a considerably
distinct behavior, due to their columnar Néel texture\citep{MoreauLuchaire2016},
sensitivity to interlayer and interfacial magnetic interactions absent
in helimagnets\citep{Finocchio2016}, and marked inhomogeneity in
their individual and collective entity due to inherent magnetic granularity\citep{Woo2016}.
Moreover, the stability of skyrmions over a wide range of thermodynamic
parameters in multilayers\citep{Soumyanarayanan2017} would enable
unprecedented investigations of topological spin excitations, including
their thermal stability and sensitivity to magnetic interactions.

\paragraph{Results Summary}

Here we present a detailed study of the magnetization dynamics of
RT Néel skyrmions in Ir/Fe/Co/Pt multilayers. A robust excitation
mode is found in ferromagnetic resonance (FMR) spectroscopy measurements,
arising from gyrotropic motion of individual Néel skyrmions, and persisting
over a wide range of temperatures and sample compositions. Micromagnetic
simulations and analytical calculations demonstrate the roles of interfacial
and interlayer interactions in shaping the distinctive resonance spectrum
of skyrmions, and enabling its detection across the granular magnetic
landscape. %

\noindent 
\section{Magnetization Dynamics\label{sec:MagDamping}}

\begin{figure}[t]
\begin{centering}
\includegraphics[width=8cm]{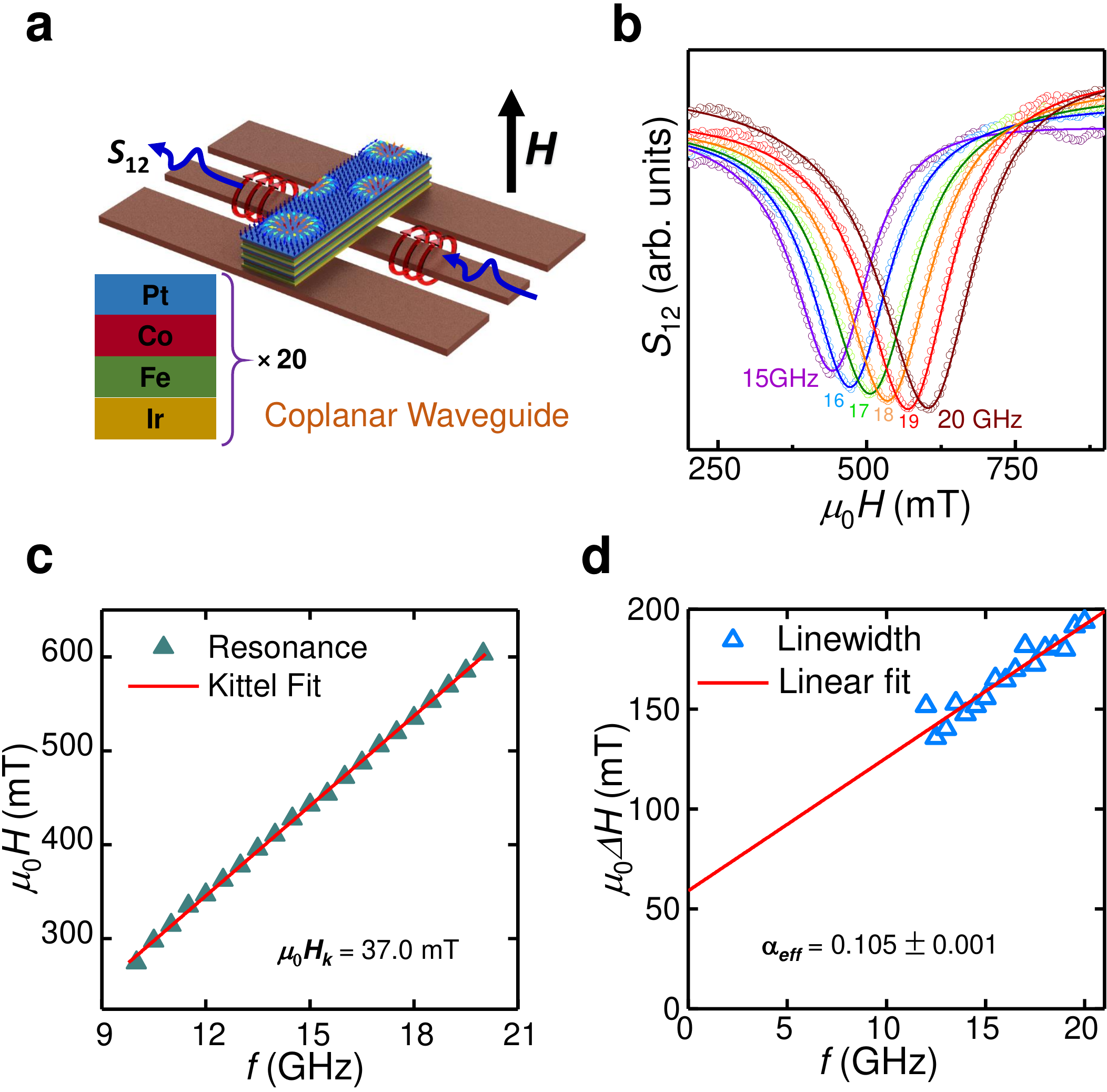}
\par\end{centering}
\noindent \caption[Magnetization Dynamics \& Damping]{\textbf{Magnetization Dynamics. (a) }Schematic of the FMR setup. The
coplanar waveguide (CPW, brown) is used to measure the transmitted
signal ($S_{12}$, blue) in response to an in-plane (IP) excitation
field ($h_{{\rm rf}}$, red) at frequency $f$ with a static out-of-plane
(OP) field $H$. Inset: schematic of the Ir/Fe/Co/Pt multilayer stack.
\textbf{(b)} FMR spectra above saturation ($H>H_{{\rm S}}$) for sample
Fe(4)/Co(6) at several frequencies between 15-20~GHz, corresponding
to uniform FM precession\citep{Kittel1948}. \textbf{(c)} Dispersion
of resonance field with frequency, determined from the Lorentzian
fits in (b), and used to extract the anisotropy field, $\mu_{0}H_{{\rm K}}$.\textbf{
(d)} Resonance linewidth $\Delta H$ plotted against frequency, and
used to determine the effective damping, $\alpha_{{\rm eff}}$.\label{fig:FMR-Damping}}
\end{figure}

\paragraph{Tunable Sk Platform}

\noindent The sputtered {[}Ir(10)/\textbf{Fe($x$)/Co($y$)}/Pt(10){]}$_{20}$
multilayer films (thickness in angstroms in parentheses) studied in
this work are known to host RT Néel skyrmions with smoothly tunable
properties that can be modulated by the Fe/Co composition\citep{Soumyanarayanan2017}.
Here we focus on two 1~nm Fe$(x)$/Co$(y)$ compositions \textendash{}
Fe(4)/Co(6) and Fe(5)/Co(5) \textendash{} which show typical skyrmion
densities $\sim30$~\textgreek{m}m$^{-2}$ (i.e. mean distance $a_{{\rm Sk}}\sim200$~nm
c.f. typical diameter, $d_{{\rm Sk}}\sim50$~nm, see §SI 1). Notably,
the large magnitude of DMI ($D\sim2.0$~mJ/m$^{2}$) relative to
the exchange stiffness ($A\sim11$~pJ/m) and out-of-plane (OP) anisotropy
($K_{{\rm eff}}\sim0.0-0.05$~MJ/m$^{3}$) results in a definitive
Néel texture (see §SI 3). Recent studies have further evidenced the
persistence of their texture and stability over a large temperature
range\citep{Yagil2017}.

\paragraph{FM Spectra \& Damping}

Broadband FMR spectroscopy measurements were performed using a home-built
coplanar waveguide (CPW) setup tailor-made for ultrathin magnetic
films (\ref{fig:FMR-Damping}a, see \nameref{sec:Methods}) \citep{Okada2017}.
FMR spectra were recorded in transmission mode ($S_{12}$) over a
frequency ($f$) range of 5-20~GHz, with the OP magnetic field ($H$)
swept over $\pm1$~T, following established recipes for multilayer
films\citep{Shaw2012,Okada2017,Turgut2017,Montoya2017}. \ref{fig:FMR-Damping}b
shows representative spectra recorded above saturation ($H>H_{{\rm S}}$):
the resonant lineshape corresponding to uniform (Kittel) precession
of the saturated ferromagnetic (FM) moment, $M_{{\rm S}}$. Lorentzian
fits to these spectra give the resonance field-frequency dispersion
(\ref{fig:FMR-Damping}c) and the linewidth (\ref{fig:FMR-Damping}d),
which are used to extract the anisotropy field, $\mu_{0}H_{{\rm K}}$\citep{Kittel1948},
and the effective damping, $\alpha_{{\rm eff}}$, respectively (see
\nameref{sec:Methods}). The considerable magnitude of $\alpha_{{\rm eff}}$
($0.105\pm0.001$) in our multilayer stack, compared to bulk crystals
and thin films, is due to the summation of various Gilbert-like damping
effects\citep{Tserkovnyak2002}, as well as interfacial effects that
include spin wave scattering, spin pumping, and interface defect scattering\citep{Shaw2012,Okada2017,Tserkovnyak2002}.
The experimental determination of $\alpha_{{\rm eff}}$ for this multilayer
skyrmion host is a crucial step towards modeling skyrmion dynamics
and switching characteristics\citep{Sampaio2013,Finocchio2016}.

\section[Microwave Resonance Phenomenology]{Resonant Magnetic Textures\label{sec:ResExcitations}}

\begin{figure}[t]
\begin{centering}
\includegraphics[width=8cm]{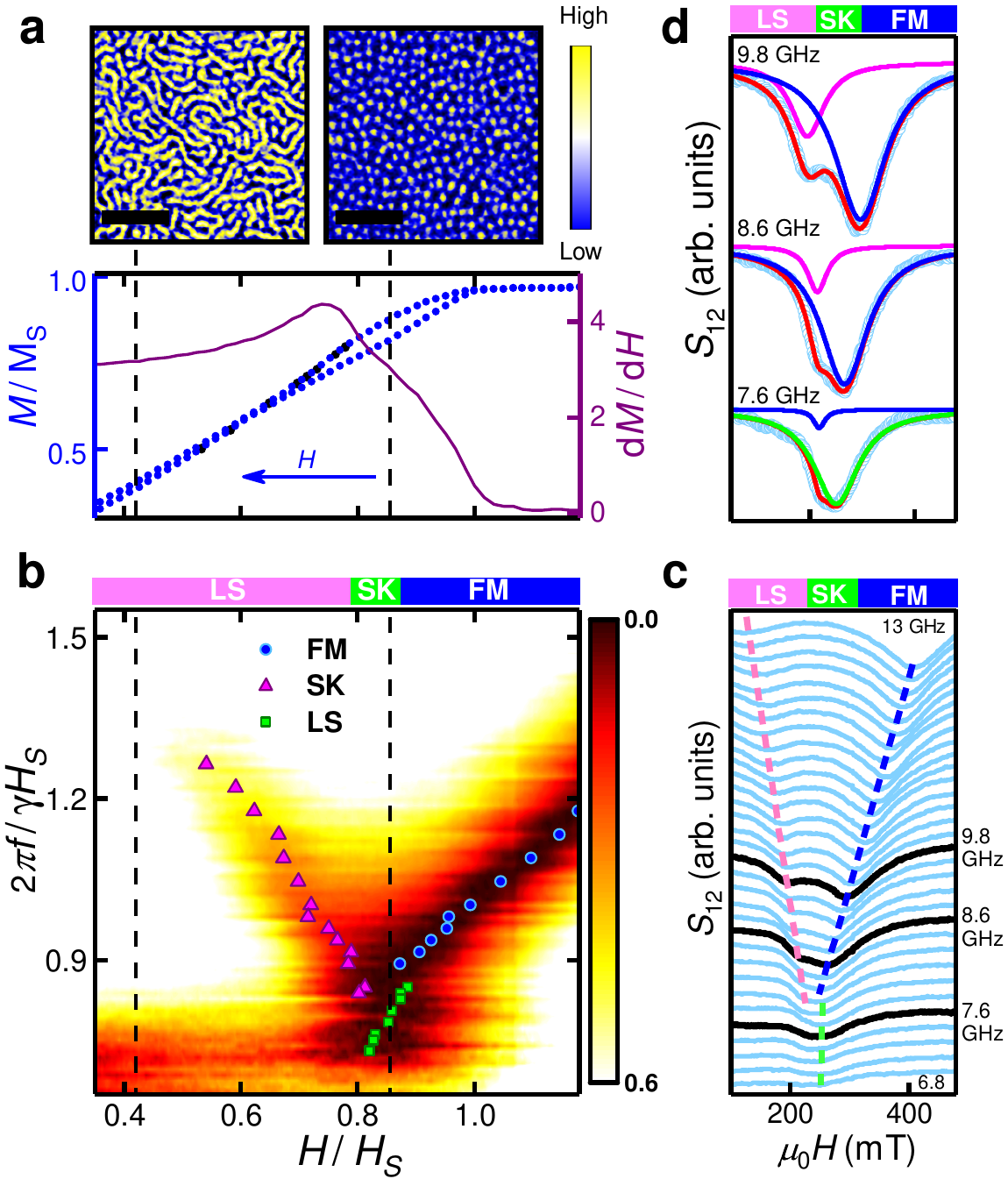}
\par\end{centering}
\noindent \caption[Observed Microwave Resonances]{\textbf{Microwave Resonances of Magnetic Textures. (a) }Normalized
OP magnetization $M/M_{{\rm S}}$ (blue) and $dM/dH$ (purple) against
$H/H_{{\rm S}}$ ($H$ swept to zero) for Fe(5)/Co(5) at RT. Top:
MFM images (scale bar: 500~nm) at fields indicated by dashed black
lines, showing labyrinthine stripe (LS, left) and skyrmion (SK, right)
phases. \textbf{(b) }Color plot of normalized FMR spectra plotted
against reduced field ($H/H_{{\rm S}}$) and frequency ($2\pi f/\gamma H_{{\rm S}}$,
$\gamma$ is the gyromagnetic ratio). Overlaid points show resonances
corresponding to three distinct modes, identified by Lorentzian fitting
(see (d)). Top: horizontal bar delineates approximate extents of LS,
SK, and FM phases. \textbf{(c)} Representative linecuts showing raw
FMR spectra at selected frequencies. Dashed lines show the dispersion
of observed minima, corresponding to resonances in (b). \textbf{(d)}
Sample Lorentizan superposition fits to FMR spectra, used to determine
the resonances (points: raw data, lines: fits). \label{fig:SkRes_Expt}}
\end{figure}

\paragraph{Magnetic Textures \& FMR Spectra}

\noindent We now turn to the microwave response of magnetic textures
that emerge at fields below saturation ($H<H_{{\rm S}}$), and form
disordered skyrmion lattice (SK, \ref{fig:SkRes_Expt}a: right inset)
and labyrinthine stripe (LS, \ref{fig:SkRes_Expt}a: left inset) phases
respectively. \ref{fig:SkRes_Expt}c shows a waterfall plot of representative
high-resolution FMR spectra which exhibit a visible dichotomy in resonance
characteristics. While high frequency spectra (e.g. 13~GHz) exhibit
two prominent minima at high ($H>H_{{\rm S}}$) and low ($H\ll H_{{\rm S}}$)
fields respectively, low frequency spectra have only one observable
dip at intermediate fields ($H\sim0.8-1\,H_{{\rm S}}$). Notably,
these minima correspond to distinctly dispersing resonances that together
form a characteristic ``Y''-shape (\ref{fig:SkRes_Expt}c). We use
Lorentzian superposition fits to determine and delineate these resonances
(e.g. \ref{fig:SkRes_Expt}d). The dispersion of the fitted resonances,
overlaid in \ref{fig:SkRes_Expt}b on a normalized spectral plot,
is used to establish their relationship with magnetic phases (\ref{fig:SkRes_Expt}a).

\paragraph{Expt Resonance Modes}

The resonances in \ref{fig:SkRes_Expt}b can be delineated into three
distinct modes \textendash{} each corresponding to a unique magnetic
phase. First, the FM phase ($H>H_{{\rm S}}$) exhibits the familiar
positively dispersing Kittel mode (blue circles) arising from uniform
precession\citep{Kittel1948}. Next, the LS phase ($H\ll H_{{\rm S}}$)
harbors a high frequency mode (magenta triangles) with negative dispersion,
analogous to well-studied helimagnonic resonances in Bloch-textured
compounds\citep{Schwarze2015}. Finally, the SK phase ($H/H_{{\rm S}}\sim0.8-1$)
hosts a distinct, low frequency resonance mode (green squares) with
positive dispersion and considerable spectral weight. This is reminiscent
of gyrotropic excitations reported in Bloch-textured magnets\citep{Onose2012,Schwarze2015,Garst2017,Turgut2017},
and can indeed be identified as a magnon-skyrmion bound state arising
from counter-clockwise (CCW) gyration of individual Néel skyrmions
as we show below. Surprisingly, this SK resonance exhibits a pronounced
renormalization towards lower frequencies with respect to the Kittel
mode, a phenomenon hitherto unobserved in any skyrmion material. %
\begin{figure}[t]
\begin{centering}
\includegraphics[width=8cm]{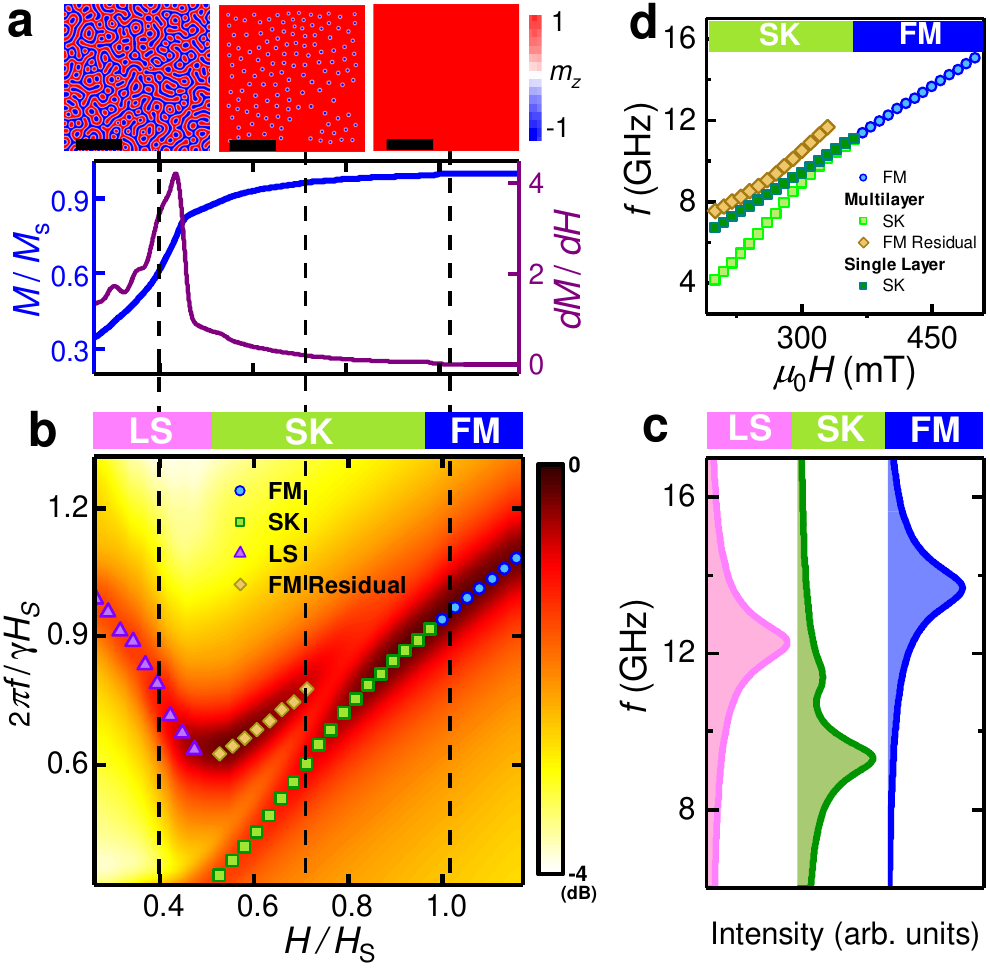}
\par\end{centering}
\noindent \caption[Simulated Microwave Resonances]{\textbf{Micromagnetic Simulation of Microwave Resonances. (a) }Simulated
equilibrium $M/M_{{\rm S}}$ (blue), and $dM/dH$ (purple) against
$H/H_{{\rm S}}$ for Fe(5)/Co(5) at RT (c.f. \ref{fig:SkRes_Expt},
details in \nameref{sec:Methods}). Top: representative OP magnetization
($m_{z}$) images (scale bar: 500~nm) at fields indicated by dashed
black lines for LS (left), SK (centre), and FM (right) phases. \textbf{(b)
}Color plot of simulated microwave resonance intensity in reduced
$f,\,H$ units (c.f. \ref{fig:SkRes_Expt}b). Overlaid points show
dispersing modes within LS, SK, and FM phases. Top: horizontal bar
delineates approximate extents of these phases. \textbf{(c)} Linecuts
extracted at dashed black lines in (b) showing typical spectra from
the respective phases. \textbf{(d)} Simulated dispersion comparison
for the multilayer (b) with a single layer with similar magnetic parameters.
The steepening of the SK mode, unique to multilayers, requires interlayer
dipolar coupling.\label{fig:SkRes_MuMag}}
\end{figure}

\paragraph{Simulated Resonance Modes}

Systematic multilayer micromagnetic simulations were performed to
establish the relationship between the magnetic textures and the observed
resonances (details in \nameref{sec:Methods}). The equilibrium magnetic
configurations exhibit consistently Néel texture (see §SI 3) with
the expected field evolution (FM, SK, and LS: \ref{fig:SkRes_MuMag}a).
Their magnetization dynamics were simulated from their temporal response
to in-plane (IP) excitation fields (\ref{fig:SkRes_MuMag}c), yielding
a spectral diagram (\ref{fig:SkRes_MuMag}b) with varying $f$ and
$H$. In agreement with experiments (\ref{fig:SkRes_Expt}b), the
resonances in FM and LS phases are at higher frequencies, and disperse
with positive and negative slopes respectively. Meanwhile, the SK
phase exhibits two positively dispersing resonances: one, at higher
frequencies (yellow diamonds), appears similar to the FM mode; while
the other, at lower frequencies (green squares), is noticeably steeper,
consistent with experiment (\ref{fig:SkRes_Expt}b). %

\noindent 
\section[Gyrotropic Skyrmion Resonances]{Gyrotropic Skyrmion Resonance\label{sec:CCW-CharDisc}}

\begin{figure}[t]
\begin{centering}
\includegraphics[width=8cm]{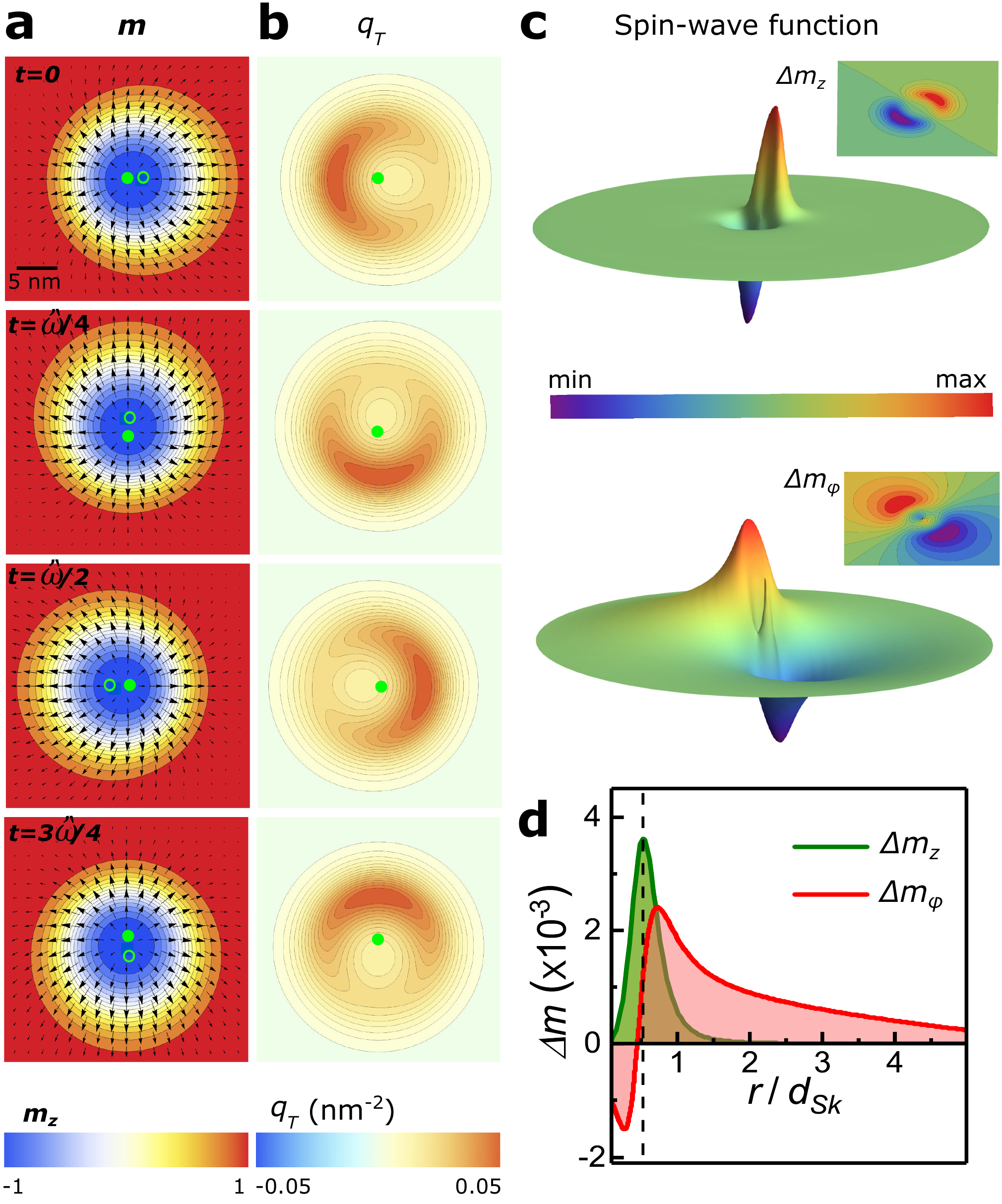}
\par\end{centering}
\noindent \caption[CCW Origin of Skyrmion Resonance]{\textbf{Gyrotropic CCW Origin of Skyrmion Resonance.} \textbf{(a-b)}
Temporal snapshots of normalized skyrmion magnetization $\boldsymbol{m}$
(a: arrows/color-scale denote IP/OP components) and its topological
charge density $q_{{\rm T}}$ (b, defined in text), showing CCW rotation
through one period ($\tau$). Filled and open green circles denote
the center-of-mass of $q_{{\rm T}}$ and $m_{z}$ respectively.\textbf{
(c)} Spin-wave function (SWF) of a resonating skyrmion at a given
instant, shown as the deviation, $\Delta\boldsymbol{m}$, from the
static configuration ($m_{z}$: OP, $m_{\varphi}$: IP), using Fourier
analysis of simulations. \textbf{(d)} Radial cuts through the maxima
of $\Delta m_{z}$ and $\Delta m_{\varphi}$, the latter possessing
a noticeably long tail. Vertical dashed line shows the skyrmion radius,
$d_{{\rm Sk}}/2$ for comparison. \label{fig:CCWTheory}}
\end{figure}

\paragraph{Theoretical Motivation}

\noindent The robust nature of the observed SK resonance (\ref{fig:SkRes_Expt}b),
characteristic of phase coherence, is particularly compelling given
that it arises from a dilute, disordered ensemble of inhomogenous
skyrmions ($\sim25\%$ size variation, see §SI 1). Furthermore, the
qualitative agreement of its distinguished phenomenology with multilayer
simulations (\ref{fig:SkRes_MuMag}b) raises several intriguing questions:
(1) what is the microscopic mechanism for the low-frequency SK resonance?
(2) what are its quantum numbers? (3) what causes the idiosyncratic
dispersion of the SK mode? We turn to spatiotemporal Fourier analyses
of simulations and analytical calculations to address these compelling
issues.

\paragraph{CCW Snapshots}

First, we examine temporal snapshots of the normalized magnetization,
$\boldsymbol{m}=\boldsymbol{M}/M_{{\rm S}}$ (\ref{fig:CCWTheory}a),
and topological charge density, $q_{{\rm T}}=\frac{1}{4\pi}\boldsymbol{m}\,(\partial_{x}\boldsymbol{m}\times\partial_{y}\boldsymbol{m})$
(\ref{fig:CCWTheory}b), of a resonant skyrmion (see \nameref{sec:Methods}).
Due to the momentum conservation\citep{Papanicolaou1991}, the center-of-mass
of $q_{{\rm T}}$, $\ensuremath{\boldsymbol{R}_{q_{{\rm T}}}}$ (\ref{fig:CCWTheory}b,
filled circle) remains static, however its maximum demonstrates CCW
rotation. Notably, the center-of-mass of $m_{z}$, $\ensuremath{\boldsymbol{R}_{m_{z}}}$
(\ref{fig:CCWTheory}a, open circle) is not static and also rotates
in a CCW direction. In this case, as in more involved types of skyrmion
dynamics, $\ensuremath{\boldsymbol{R}_{q_{{\rm T}}}}$ and $\ensuremath{\boldsymbol{R}_{m_{z}}}$
behave like massless and massive particles respectively\citep{Kravchuk2017,Buttner2015}.
These analyses (details in §SI 5) conclusively inform that the observed
SK resonance arises from the gyrotropic CCW excitation of single Néel
skyrmions.

\paragraph{Spin-Wave Function \& Coupling}

Second, for a dilute ensemble as here (typical $a_{{\rm Sk}}/d_{{\rm Sk}}\sim4\gg1$),
skyrmions should respond individually to the excitation field, in
contrast to collective excitations of well-studied helimagnetic skyrmion
lattices\citep{Schwarze2015,Garst2017}. It is noteworthy that the
development of collective coherence is not essential for experimental
detection: the incoherent response of a finite density of skyrmions
could produce a large enough signal to explain their observed signature.
Such an excitation can therefore be described by a \textbf{\emph{well-localized
spin-wave function (SWF)}}. The SWF (\ref{fig:CCWTheory}c, §SI 6),
determined from the temporal deviations in magnetization $\Delta\boldsymbol{m}$
from the static skyrmion configuration, displays profiles characteristic
of an individual CCW excitation with appropriate quantum numbers (details
in \nameref{sec:Methods}). Next, radial linecuts through SWF maxima
for OP ($\Delta m_{z}$) and IP ($\Delta m_{\varphi}$) components
(\ref{fig:CCWTheory}d) peak as expected at the skyrmion radius ($d_{{\rm Sk}}/2$).
In contrast to the single-layer theory (§SI 5), the long-range dipolar
field in the multilayer structure results in a long tail in $\Delta m_{\varphi}$.
Interestingly, this long tail might give rise to a residual coupling
between adjacent skyrmions that could lead to a synchronization of
excitations, a subject for future investigation.

\paragraph{Renormalized Dispersion}

Third, we investigate the renormalization of the multilayer CCW mode
to lower frequencies with respect to the Kittel mode. Analytical calculations
of the Néel skyrmion spectrum for a comparable \textbf{\emph{single
layer}} (§SI 5)\citep{Sheka2001} suggest that the CCW mode can correspond
only to a shallow bound state, with frequency practically indistinguishable
from the Kittel mode, as for Bloch systems\citep{Lin2014,Schutte2014}.
Indeed, this is verified by single layer simulations (\ref{fig:SkRes_MuMag}d,
§SI 4), which do not exhibit renormalization, in line with literature\citep{Zhang2016,Guslienko2017}.
Together, these conclusively point to interlayer dipolar coupling,
uniquely present in multilayers, as being responsible for the CCW
renormalization. The dipolar coupling aligns layer-wise moments within
multilayer skyrmions, producing an attractive interaction that may
lower the gyrotropic frequency well below the magnon continuum. Meanwhile,
the higher frequency SK mode, uniquely observed in simulations (\ref{fig:SkRes_MuMag}b:
FM residual), arises from a delocalized excitation of the background,
reminiscent of the uniform Kittel mode\citep{Kittel1948}. It is spectrally
distinguished in multilayers due to the stronger dispersion of the
CCW, while its substantially lower weight is likely the reason for
its absence within experimental resolution. %
\begin{figure}[t]
\begin{centering}
\includegraphics[width=8cm]{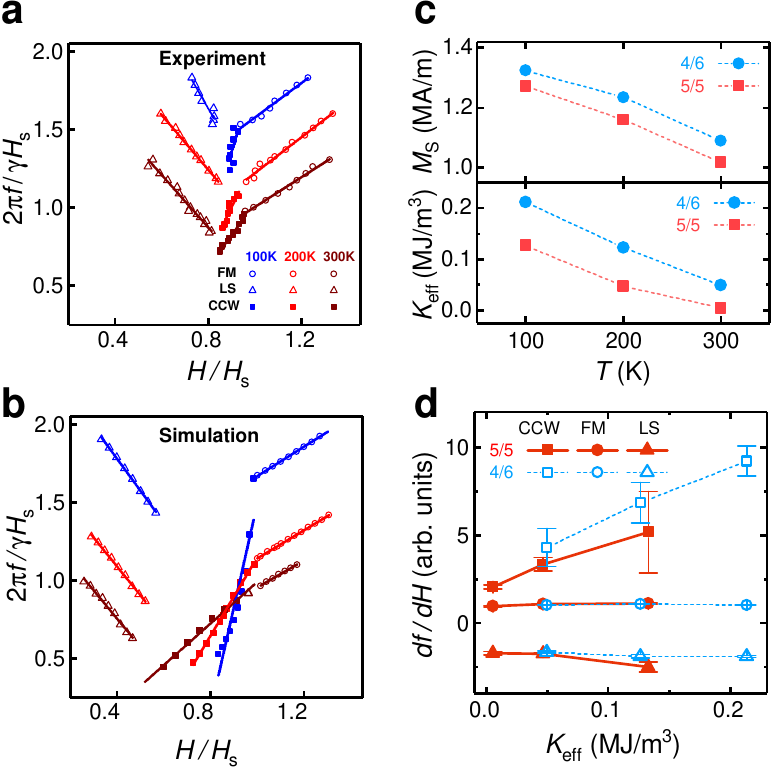}
\par\end{centering}
\noindent \caption[Thermodynamic Evolution of Resonances]{\textbf{Thermodynamic Evolution of Resonances. (a-b) }Dispersions
of LS, SK (CCW), and FM resonance modes for Fe(5)/Co(5) at selected
temperatures ($T=$100, 200, 300 K) from (a) experiment, and (b) simulations
(multilayers, 0~K with $T$-dependent parameters, see \nameref{sec:Methods}).
\textbf{(c)} $T$-dependence of experimentally determined values for
saturation magnetization $M_{{\rm S}}$ (top) and $K_{{\rm eff}}$
(bottom) for samples Fe(5)/Co(5) and Fe(4)/Co(6).\textbf{ (d)} Measured
resonance dispersions, $df/dH$ over 100-300~K for both samples,
obtained from linear fits (e.g. to a), plotted against $K_{{\rm eff}}$.
\label{fig:SkRes_TDep} }
\end{figure}

\paragraph{T-Dependence}

Finally, the persistence of Ir/Fe/Co/Pt skyrmions over a wide range
of thermodynamic parameters\citep{Soumyanarayanan2017,Yagil2017}
offers an unprecedented window into the evolution of skyrmion resonances.
To this end, we examine the temperature variation (100-300~K) of
Fe(5)/Co(5) resonances within experiments (\ref{fig:SkRes_TDep}a)
and simulations (\ref{fig:SkRes_TDep}b, parameters in \nameref{sec:Methods})
respectively. The FM and LS modes do not vary substantially, however
the CCW mode is noticeably steeper at lower temperatures (\ref{fig:SkRes_TDep}a).
While zero Kelvin simulations cannot establish full quantitative consistency
with experiment, they do reproduce the CCW trend (\ref{fig:SkRes_TDep}b),
thereby explicitly discounting thermal activation effects. Therefore,
we investigate its thermodynamic origin by studying the evolution
of resonance mode dispersions, ${\rm d}f/{\rm d}H$, across two samples
over these temperatures. From \ref{fig:SkRes_TDep}d, the CCW dispersion
steepens by a factor of 5, and notably exhibits a near-linear variation
with $K_{{\rm eff}}$ \textendash{} the latter also increases considerably
(0.01\textendash 0.2~MJ/m$^{3}$) across our measurements. Indeed,
theoretical reports point to easy-axis anisotropy favouring the gyrotropic
CCW mode and lowering its resonance frequency\citep{Sheka2001,Kravchuk2017},
consistent with our findings. %

\section{Outlook\label{sec:Outlook}}

\paragraph{Conclusions}

\noindent We have reported a systematic magnetization dynamics study
of Néel skyrmions that form dilute, inhomogeneous ensembles in Ir/Fe/Co/Pt
multilayers. We have identified a strong microwave resonance arising
from CCW gyration of individual skyrmions, and note its persistence
over a wide range of thermodynamic parameters. Multilayer simulations
and analytical calculations unveil the roles of interlayer dipolar
coupling and easy-axis anisotropy in renormalizing the skyrmion resonance
spectrum against the magnon continuum. These results establish a comprehensive
foundational scaffold between multilayer magnetic interactions and
magnetization dynamics of Néel skyrmions, and provide quantifiable
insights on their stability and dynamics.

\paragraph{Future Directions}

The excitation of RT skyrmions in multilayers is remarkable in light
of inherent skyrmionic granularity and inhomogeneity, and is promising
towards their exploitation for new physics and technology. Such excitations
enable deterministic nucleation and manipulation of skyrmions in device
configurations\citep{Finocchio2016}, and can be used to realize skyrmion-based
microwave detectors and oscillators\citep{Finocchio2015,Garcia-Sanchez2016}.
The renormalization of the CCW mode due to the interlayer coupling
clearly separates this resonance from the continuum, which is key
for an interference-free functionalization involving skyrmion-based
data transmission\citep{Soumyanarayanan2016} and magnon-based logic\citep{Chumak2015}
in a single device. Our work forms a cornerstone for mechanistic tailoring
of skyrmion dynamics in technologically relevant systems\citep{Finocchio2016},
and opens a new chapter on topological magnonics\citep{Chumak2015,Garst2017}.

\noindent \begin{center}
\rule[0.5ex]{0.6\columnwidth}{1pt}
\par\end{center}

\noindent \textbf{\emph{\small{}Acknowledgments.}}\textbf{\small{}
}{\small{}We acknowledge Anthony Tan and Ophir Auslaender for experimental
inputs, and Ulrike Nitzsche for technical assistance. We also acknowledge
the support of the National Supercomputing Centre (NSCC), Singapore,
the A{*}STAR Computational Resource Center (A{*}CRC), Singapore for
computational work. This work was supported by the the A{*}STAR Pharos
Fund of Singapore (Ref. No. 1527400026), the Ministry of Education
(MoE), Academic Research Fund Tier 2 (Ref. No. MOE2014-T2-1-050) of
Singapore, and the National Research Foundation (NRF) of Singapore,
NRF - Investigatorship (Ref. No.: NRF-NRFI2015-04). V.K. acknowledges
support from the Alexander von Humboldt Foundation and the National
Academy of Sciences of Ukraine (Project No. 0116U003192). M.G. acknowledges
financial support from DFG CRC 1143 and DFG Grant 1072/5-1.}{\small \par}

\noindent \textbf{\emph{\small{}Author Contributions.}}\textbf{\small{}
}{\small{}A.S. and C.P. conceived the research and coordinated the
project. M.R. deposited and characterized the films with A.S. S.H.
and C.P. designed the FMR experiments. B.S. and S.H. performed the
experiments, and F.M., B.S., and S.H. analyzed the data. F.M. performed
the micromagnetic simulations. V.K. and M.G. performed the theoretical
calculations. All authors discussed the results and contributed to
the manuscript.}{\small \par}

\bibliographystyle{apsrev4-1}
\bibliography{SkFMR}
\begin{center}
\rule{0.6\columnwidth}{1pt}
\par\end{center}

\renewcommand\thefigure{M\arabic{section}}\renewcommand\thetable{M\arabic{table}}\renewcommand\theequation{M\arabic{equation}}
\begin{small}

\section{Methods\label{sec:Methods}}

\subsection{Film Deposition \& Characterization\label{par:FilmDep-Methods}}

\noindent Multilayer films of: \\
Ta(30)/Pt(100)/{[}Ir(10)/\textbf{Fe($x$)/Co($y$)}/Pt(10){]}$_{20}$/Pt(20)
\\
(numbers in parentheses denote nominal layer thickness in angstroms)
were sputtered on thermally oxidized 100~mm Si wafers. The deposition
parameters, and structural and magnetic properties of the multilayer
stacks have been reported previously\citep{Soumyanarayanan2017}. 

The results reported here correspond to two compositions: Fe(4)/Co(6)
and Fe(5)/Co(5). Magnetization $M(H)$ were measured using a Quantum
Design\texttrademark{} Magnetic Properties Measurement System (MPMS)
in OP/IP configuration to determine $H_{{\rm S}}$, $M_{{\rm S}}$,
and $K_{{\rm eff}}$ (§SI 1). MFM images were acquired using a D3100
AFM manufactured by Bruker\texttrademark , mounted on a vibration-isolated
platform, using sharp, ultra-low moment SSS-MFMR\texttrademark{} tips.
The reported skyrmion properties are consistent with previous RT results
(see §SI 1)\citep{Soumyanarayanan2017}, recently extended to temperatures
down to 5~K\citep{Yagil2017}. 

\subsection{FMR Measurements\label{par:FMR-Methods}}

\noindent Broadband FMR measurements were performed using a home-built
setup\citep{He2016}, previously used for high-resolution spectroscopy
of ultrathin magnetic films\citep{Okada2017}. A 4\texttimes 8~mm
sample was inductively coupled to a U-shaped CPW using a spring-loaded
sample holder. Microwave excitations were sourced parallel to the
sample plane (\ref{fig:FMR-Damping}a), and the transmitted signal
($S_{12}$) was measured using a Keysight PNA N5222 vector network
analyzer (VNA). External OP magnetic fields up to \textpm 1~T were
applied to saturate the sample, and data were recorded at each frequency
in field sweep mode. The results presented here correspond to $H$
swept from above $+H_{{\rm S}}$ to zero (full sweep in §SI 2). Low
temperature measurements were performed with the sample holder mounted
on a stage attached to a continuous flow cryostat\citep{Okada2017}. 

For uniform Kittel precession ($H>H_{{\rm S}}$) with frequency $f_{{\rm r}}$,
the resonance field $H_{{\rm r}}$ and linewidth $\Delta H_{{\rm r}}$
were extracted by fitting the spectra to the expected form of the
OP dynamic susceptibility\citep{Okada2017}. The $H_{{\rm r}}-f_{{\rm r}}$
dispersion of the uniform mode (\ref{fig:FMR-Damping}c) can be described
by the OP Kittel equation\citep{Kittel1948}: 
\begin{equation}
2\pi f_{{\rm r}}=\gamma(\mu_{0}H_{{\rm r}})-\gamma\mu_{0}(H_{{\rm K}}-M_{{\rm S}})\label{eq:Kittel_OPRes}
\end{equation}
The gyromagnetic ratio $\gamma=g\mu_{{\rm B}}/\hbar$ and the anisotropy
field, $\mu_{0}H_{{\rm K}}$ can be extracted from \ref{eq:Kittel_OPRes}
(\ref{fig:FMR-Damping}c, see §SI 1). Meanwhile, the linewidth $\Delta H_{{\rm r}}$
(\ref{fig:FMR-Damping}d) is described by\citep{Shaw2012}:
\begin{equation}
\Delta H_{{\rm r}}=\Delta H_{0}+\frac{4\pi\alpha_{{\rm eff}}}{\mu_{0}\gamma}\,f_{{\rm r}}\label{eq:Damping_Form}
\end{equation}
A linear fit to $\Delta H_{{\rm r}}-f_{{\rm r}}$ (\ref{fig:FMR-Damping}d)
determines the inhomogeneous broadening, $\Delta H_{0}$, and importantly,
the effective damping, $\alpha_{{\rm eff}}$ \citep{Okada2017}.

For data acquired at or below $H_{{\rm S}}$ (\ref{fig:SkRes_Expt}),
the resonances modes for LS, SK, and FM phases were determined by
fitting the spectra to a superposition of Lorentzian lineshapes. Typical
fit results are shown in \ref{fig:SkRes_Expt}b. The spectra were
plotted with reduced field ($H/H_{{\rm S}}$) and frequency ($2\pi f/\gamma H_{{\rm S}}$)
units (\ref{fig:SkRes_Expt}b) for comparison with simulations (\ref{fig:SkRes_MuMag}b),
and across samples (\ref{fig:SkRes_TDep}). 

\subsection{Micromagnetic Simulations\label{par:MuMag-SimParams}}

\noindent Micromagnetic simulations of equilibrium magnetic textures
and their microwave response were performed using the mumax\textthreesuperior{}
package\citep{Vansteenkiste2014}, which accounts for interfacial
DMI. The multilayers were modeled with a mesh size of 4 \texttimes{}
4 \texttimes{} 1~nm over a 2 \texttimes{} 2~\textgreek{m}m area
for comparison with experiments. The magnetic (Fe, Co) layers were
modeled as a 1~nm FM layer, while the non-magnetic (Pt, Ir) layers
were represented as 1~nm spacers. The magnetic parameters $M_{{\rm S}}$
and $K_{{\rm eff}}$ were determined from SQUID magnetometry, $\alpha$
from FMR measurements, and $A$ and $D$ from micromagnetic fits to
MFM measurements using established techniques\citep{MoreauLuchaire2016,Woo2016}.
The RT parameters have been reported previously\citep{Soumyanarayanan2017},
and those for lower temperatures were determined similarly (see \ref{fig:SkRes_TDep}c,
§SI 3). Notably, all simulations were performed at 0~K.

First, a set of simulations were performed using the 20 stack repeats
configuration consistent with experiment (\textbf{\emph{Full Stack}}).
The magnetization was randomized at $H=0$, and relaxed to obtain
the equilibrium configuration. Then $H$ was increased progressively
up to $H_{{\rm S}}$ to simulate the field evolution of magnetic textures.
Layer-wise analyses of spin configurations in the SK phase reveal
uniformly Néel-textured skyrmions with size variations of up to 10\%
across layers (detailed analysis in §SI 3). The consistency of spin
configuration across layers is in line with the large magnitude of
DMI ($\sim$2~mJ/m$^{2}$) in our stacks. 

Subsequently, the simulations were repeated using periodic boundary
conditions (\textbf{\emph{PBC}}) in the $z$ direction (9 repeats
each along $\pm z$) to mimic the Full Stack behavior, i.e. to incorporate
interlayer dipolar coupling without the additional $\sim20\times$
computational cost. The PBC results were consistent with Full Stack
(within $\sim10\%$, see §SI 3), and with magnetometry and MFM experiments.
Computationally expensive magnetization dynamics work followed a similar
protocol: quantitative comparison of Full Stack and PBC for one set
of parameters (Fe(5)/Co(5) at 300~K, see §SI 4), followed by PBC
for the remainder. 

The magnetization dynamics were simulated by examining the response
of the equilibrium magnetic textures to a spatially uniform IP excitation
field (see §SI 4), 
\begin{equation}
h(t)=h_{0}\,\sin\left(2\pi f_{{\rm c}}t\right)/\left(2\pi f_{{\rm c}}t\right)\label{eq:mumag_ACexc}
\end{equation}
Here, we set $h_{0}$ to 10~mT, and $f_{{\rm c}}$ to 50~GHz \textendash{}
the latter being above the experimental bandwidth. The transient dynamics
were computed over 10~ns, and the power spectral density was determined
from the Fourier transform of the spatially averaged magnetization
response. This procedure was repeated at all relevant $H$ values
to obtain \ref{fig:SkRes_MuMag}b. Systematics associated with damping,
grid size, and interlayer interactions are detailed in §SI 4. 

\subsection{Fourier Analysis of Skyrmion Resonances\label{par:FourierAnalysis-Sims}}

\noindent The skyrmion excitations at a given $(H,\,f)$ was visualized
applying a sinusoidal IP excitation field, $h(t)=h_{0}\sin\left(2\pi ft\right)$,
and acquiring snapshots of the $\boldsymbol{m}(r)$ (\ref{fig:CCWTheory}a)
and $q_{{\rm T}}(r)$ (\ref{fig:CCWTheory}b) at 10 ps intervals (\ref{fig:CCWTheory}a-b).
The centres-of-mass (circles in \ref{fig:CCWTheory}a-b) are determined
for $m_{z}$ to be $R_{m_{z}}=\int dr\,r(1-m_{z})/\int dr(1-m_{z})$,
and for $q_{{\rm T}}$ to be $R_{q_{T}}=\int dr\,rq_{{\rm T}}(r)/Q_{{\rm Sk}}$,
where the net topological charge, $Q_{{\rm Sk}}=\int dr\,q_{{\rm T}}(r)=-1$
for a skyrmion. 

To analytically establish the origin of these resonances, the $\boldsymbol{m}(\boldsymbol{r})$
for each temporal snapshot was first projected onto a cylindrical
reference frame $\{\hat{\boldsymbol{e}}_{r},\hat{\boldsymbol{e}}_{\varphi},\hat{\boldsymbol{e}}_{z}\}$
centered on $\boldsymbol{R}_{q_{{\rm T}}}$, which remains static
through the period of the simulation ($\tau$). Next, the deviations
of OP ($\Delta m_{z}=m_{z}-m_{z,0}$) and IP ($\Delta m_{\varphi}=m_{\varphi}-m_{\varphi,0}$)
components relative to the static skyrmion configuration $\boldsymbol{m}_{0}$
were determined (note: $m_{\varphi,0}=0$ for a Néel skyrmion). The
spatiotemporal deviation thus obtained, $\Delta\boldsymbol{m}\,(t,r,\varphi)$
was then Fourier transformed for azimuthal separation of resonance
modes
\begin{equation}
F_{\mu}^{(z,\varphi)}(\omega)=\frac{1}{\pi\mathcal{R}^{2}\tau}\int\limits _{0}^{\tau}\mathrm{d}t\int\limits _{0}^{\mathcal{R}}\mathrm{d}r\,r\int\limits _{0}^{2\pi}\mathrm{d}\varphi\,\Delta m_{z,\varphi}(t,r,\varphi)e^{i(\omega t+\mu\varphi)},\label{eq:FT_MagDev}
\end{equation}
where $\mathcal{R}$ is radius of the disk-shaped sample and importantly
$\mu\in\mathbb{Z}$ is the azimuthal wave number. Spectra $F_{\mu}^{(z,\varphi)}(\omega)$
are calculated for $|\mu|\le4$ (i.e. nine values of $\mu$). Both
$\Delta m_{z}$ and $\Delta m_{\varphi}$ exhibit the most intense
response for $\mu=-1$ (CCW mode). The Fourier spectra, $F_{-1}^{(z,\varphi)}(\omega)$,
have a well-pronounced discrete structure for the resonances $\{\omega_{i}\}$.
The lowest of these, $\omega_{\text{ccw}}=\min_{i}\{\omega_{i}\}$,
is recognized as the eigen-frequency if the CCW mode.

To extract the radial profiles of the CCW mode, we calculate:
\begin{equation}
\mathcal{F}^{(z,\varphi)}(r)=\frac{1}{2\pi\tau}\int\limits _{0}^{\tau}\mathrm{d}t\int\limits _{0}^{2\pi}\mathrm{d}\varphi\,\Delta m_{z,\varphi}(t,r,\varphi)e^{i(\bar{\omega}t+\bar{\mu}\varphi)},\label{eq:F-G}
\end{equation}
where $\bar{\mu}=-1$, and $\bar{\omega}=\omega_{{\rm CCW}}$ is determined
above. The radial profiles $\tilde{m}_{z}(r)$ and $\tilde{m}_{\varphi}(r)$
can then be obtained by accounting for the general symmetry\citep{Kravchuk2017}
$\Delta m_{z}(t,r,\varphi)=\tilde{m}_{z}(r)\cos(\omega t+\mu\varphi+\eta)$,
and $\Delta m_{\varphi}(t,r,\varphi)=\tilde{m}_{\varphi}(r)\sin(\omega t+\mu\varphi+\eta)$,
where $\eta$ is an arbitrary phase. In this case $\mathcal{F}^{z}(r)\approx\frac{1}{2}\tilde{m}_{z}(r)e^{-i\eta}$
and $\mathcal{F}^{\varphi}(r)\approx\frac{i}{2}\tilde{m}_{\varphi}(r)e^{-i\eta}$.
Thus, one can reconstruct the CCW mode as
\begin{align}
\Delta m_{z,\varphi}(t,r,\varphi) & = & 2\cdot\text{Re}[\mathcal{F}^{(z,\varphi)}(r)]\cos(\bar{\omega}t+\bar{\mu}\varphi)\label{eq:CCW_Reconst}\\
 & + & 2\cdot\text{Im}[\mathcal{F}^{(z,\varphi)}(r)]\sin(\bar{\omega}t+\bar{\mu}\varphi)\nonumber 
\end{align}
Results of this reconstruction are shown in \ref{fig:CCWTheory}c-d.

\end{small}

\noindent 
\end{document}